\documentclass{llncs}
\usepackage{wrapfig}
\usepackage{llncsdoc}
\usepackage{epsfig,psfrag}
\usepackage{latexsym,amssymb,amsmath}
\usepackage{longtable}
\usepackage{tikz,pgf}
\usepackage{subfig}
\usepackage{wrapfig}
\usepackage{rotating}
\usepackage{algorithm}
\usepackage{algorithmic}

\usepackage{graphicx}
\usepackage{epstopdf}
\usepackage{color}

\usepackage{amsmath,amsfonts}
\newcommand{\bqn}{\begin{eqnarray}}
\newcommand{\eqn}{\end{eqnarray}}
\newcommand{\bq}{\begin{eqnarray*}}
\newcommand{\eq}{\end{eqnarray*}}

\usepackage{url}
\usepackage{hyperref}
\hypersetup{colorlinks,%
citecolor=blue,%
filecolor=blue,%
linkcolor=red,%
urlcolor=blue,%
pdftex}

\newcommand{\red}[1]{\textcolor{red}{#1}}

\begin{document}

\title{Topological Distances between Networks \\ 
and Its Application to Brain Imaging}

\author{Hyekyoung Lee\inst{1}, Zhiwei Ma \inst{2}, Yuan Wang \inst{3}, Moo K. Chung\inst{3}}
\institute{
$^1$ Seoul National University Hospital, Korea\\
$^2$University of Chicago, USA\\
$^3$University of Wisconsin-Madison, USA\\
\vspace{0.3cm}
\red{\tt hklee.brain@gmail.com}  \hspace{0.15cm} \red{\tt mkchung@wisc.edu}}

\maketitle

\pagestyle{headings}
\setcounter{page}{1}
\pagenumbering{arabic}

\begin{abstract}
This paper surveys various distance measures for networks and graphs that were introduced in  persistent homology. 
The scope of the paper is limited to network distances that were actually used in brain networks but the methods can be easily adapted to any weighted graph in other fields. The network version of Gromov-Hausdorff, bottleneck, kernel distances are introduced. We also introduce a recently developed KS-test like distance based on monotonic topology features such as the zeroth Betti number. Numerous toy examples and the result of applying many different distances to the brain networks of different clinical status and populations are given. 
\end{abstract}

\section{Introduction}

There are many statistically oriented similarity measures and distances between networks in literature \cite{banks.1994,chen.2016,Petri2014}. Many of these approaches simply ignore the topology of the networks and mainly use the sum of differences between either node or link measurements or correlations. These network distances are sensitive to the topology of networks. They may lose sensitivity over topological structures such as the connected components or holes in the network.

In graph theoretic approaches,  the similarity and distance of networks are measured by determining the difference in graph theory features such as assortativity, betweenness centrality, small-worldness and network homogeneity \cite{bullmore.2009.nrn,rubinov.2009.ni,uddin.2008}. Comparison of graph theory features appears to reveal changes of structural or functional connectivity associated with different clinical populations \cite{rubinov.2009.ni}. Since weighted brain networks are difficult to interpret and visualize, they are often turned into binary networks by thresholding edge weights \cite{he.2008,vanwijk.2010.plosone}. However, the choice of thresholding the edge weights may alter  the network topology. To obtain the proper optimal threshold, the multiple comparison correction over every possible edge has been proposed \cite{bohland.2009.plosone,chen.2008.cc,ferrarini.2009.hbm,rubinov.2009.hbm,salvador.2005.cc,vanwijk.2010.plosone}. However, depending on what $p$-value to threshold, the resulting binary graph also changes. Others tried to control the sparsity of edges in the network in obtaining the binary network \cite{achard.2007.ploscb,bassett.2006.adaptive,he.2008,kramer.2008.er,meunier.2009.ni,vanheuvel.2009.jneuro,vanwijk.2010.plosone}. But then one encounters the problem of thresholding sparse parameters. Thus existing methods for binarizing weighted networks have been circular. Until now, there are not widely accepted criteria for thresholding networks. Instead of trying to come up with a proper threshold for network construction that may not work for different clinical populations or cognitive conditions \cite{vanwijk.2010.plosone}, why not use all networks for every possible threshold? Motivated by this question, new multiscale hierarchical network modeling framework based on persistent homology has been developed recently \cite{chung.2013.MICCAI,chung.2015.TMI,lee.2011.MICCAI,lee.2011.ISBI,lee.2012.TMI}.

Persistent homology, a branch of computational topology \cite{carlsson.2008,edelsbrunner.2008}, provides a more coherent mathematical framework to measuring network distance than the conventional method of simply taking difference between graph theoretic features. Instead of looking at networks at a fixed scale, as usually done in many standard brain network analysis, persistent homology observes the changes of topological features of the network over multiple resolutions and scales \cite{edelsbrunner.2009,horak.2009.jsm,zomorodian.2005.dcg}. In doing so, it reveals the most persistent topological features that are robust under noise perturbations. This robustness in performance under different scales is needed for most network distancs that are parameter and scale dependent.

In persistent homological network analysis as established in \cite{lee.2011.MICCAI,lee.2012.TMI},  instead of analyzing networks at one fixed threshold that may not be optimal, we build the collection of nested networks over every possible threshold using a {\em graph filtration}, a persistent homological construct  \cite{chung.2013.MICCAI,lee.2012.TMI}. The graph filtration is a threshold-free framework for analyzing a family of graphs but requires hierarchically building specific nested subgraph structures. The proposed method share some similarities to the existing multi-thresholding or multi-resolution network models  that use many different arbitrary thresholds or scales  \cite{achard.2006,he.2008,kim.2015,lee.2012.TMI,Supekar2008}. Such approaches are mainly used to visually display the dynamic pattern of how graph theoretic features change over different thresholds and the pattern of change is rarely quantified. Persistent homology can be used to quantify such dynamic pattern in a more coherent quantifiable way.

In persistent homology, there are various metrics that have been proposed to measure distance between topological spaces. 
The {\em bottleneck distance} is perhaps the most widely used distance measure in persistent homology \cite{cohensteiner.2007,Cohen-Steiner2010}. The bottleneck distance was originally defined for measuring the distance between two sets in the same metric space. The method was later adapted to measure the difference between two persistence diagrams \cite{edelsbrunner.2009}. 
Despite the strong mathematical motivation from differential geometry, they are not intuitive, as points in different persistence diagrams (PD) do not naturally correspond as a bijection. Thus, the computation of the distances between PDs is not straightforward \cite{Cohen-Steiner2010,heo.2012,Lovasz1986}. \cite{Chung2009} bypassed the problem of comparing PDs by that of comparing the density of points in a squre grid {\em via} kernel smoothing. Motivated by \cite{Chung2009}, later theories and applications of inference on PDs have taken the direction of {\em kernel distances} \cite{pachauri.2011,reininghaus.2015,lee.2016.OHBM}. The kernel itself is an inner product; thus, it naturally yields a Hilbert space structure. Subsequently, other Hilberst space tools such as principle component analysis or support vector machine can be easily applied. In \cite{reininghaus.2015}, persistent scale space (PSS) kernel, which is the inner product of heat diffusions with the PD points as the initial heat sources, is introduced. PSS is later extended to multiple PSS kernels in determining the changes in brain network distances in aging \cite{lee.2016.OHBM}.

{\em Gromov-Hausdorff (GH) distance} is also another popular distance that is originally used to measure distance between two metric spaces \cite{tuzhilin.2016}. It was later adapted to measure distances in persistent homology and dendrograms \cite{carlsson.2008,carlsson.2010.jmlr,chazal.2009}. Subsequently, it was applied to brain networks \cite{lee.2011.MICCAI,lee.2012.TMI}.  In the case of brain networks, we often have prior node-to-node correspondences established by image registration. Thus, the computation of GH-distance is simplified to the differences in single linkage matrices. This connects GH-distance to popular hierarchical clustering.

All the above network distances do not have known probability distributions. The statistical inferences on distances have been done through resampling techniques such as jackknife, bootstraps or permutations \cite{lee.2012.TMI,lee.2017.HBM,chung.2015.TMI}, which often cause computational bottlenecks for large scale networks. To bypass the computational bottleneck associated with resampling large scale networks, 
the {\em Kolmogorove-Smirnov (KS) test like distance} was first proposed in \cite{chung.2013.CNA,chung.2013.MICCAI}. Later it was also used in characterizing the multidimensional graph filtration in multimodal brain imaging study \cite{lee.2017.HBM}. The advantage of using KS-test like distance is its easy to interpret compared to other less intuitive distances. Due to its simplicity, it is possible to enumerate the possible sample space combinatorially and determine its probability distribution exactly without estimating the distribution empirically by resampling techniques \cite{chung.2016.twin}.

Many distance or similarity measures are not metrics but having metric distances makes the interpretation of brain networks easier due to the triangle inequality. Many network distances are in fact metric or ultrametric \cite{lee.2012.TMI}. Let us start with the review of metric space and formulate networks as metric spaces.

\section{Network as a metric space}
\label{sec:network_construction}

Consider a weighted graph or network with the node set $V = \left\{ 1, \dots, p \right\}$ and the edge weights $\rho=(\rho_{ij})$, where $\rho_{ij}$ is the weight between nodes $i$ and $j$.
The edge weight is usually given by a similarity measure between the observed data on the nodes. Various similarity measures  have been proposed. The correlation or mutual information between measurements for the biological or metabolic network and the frequency of contact between actors for the social network have been used as edge weights \cite{bassett.2006.neuroscientist,bien.2011}.

We will assume that the edge weights satisfy the metric properties: nonnegativity, identity, symmetry and the triangle inequality such that
$$\rho_{i,j} \geq 0, \; \rho_{ii} =0\; \rho_{ij} = \rho_{ji}, \;\rho_{ij} \leq \rho_{ik} + \rho_{kj}.$$ 
With theses conditions, $\mathcal{X}=(V, \rho)$ forms a metric space. Many real-world networks satisfy the metric properties. Although the metric property is not necessary for building a network, it offers many nice mathematical properties and easier interpretation on network connectivity. 

\begin{wrapfigure}{r}{0.5\textwidth}
\vspace{-0.5cm}
\centering
\includegraphics[width=1\linewidth]{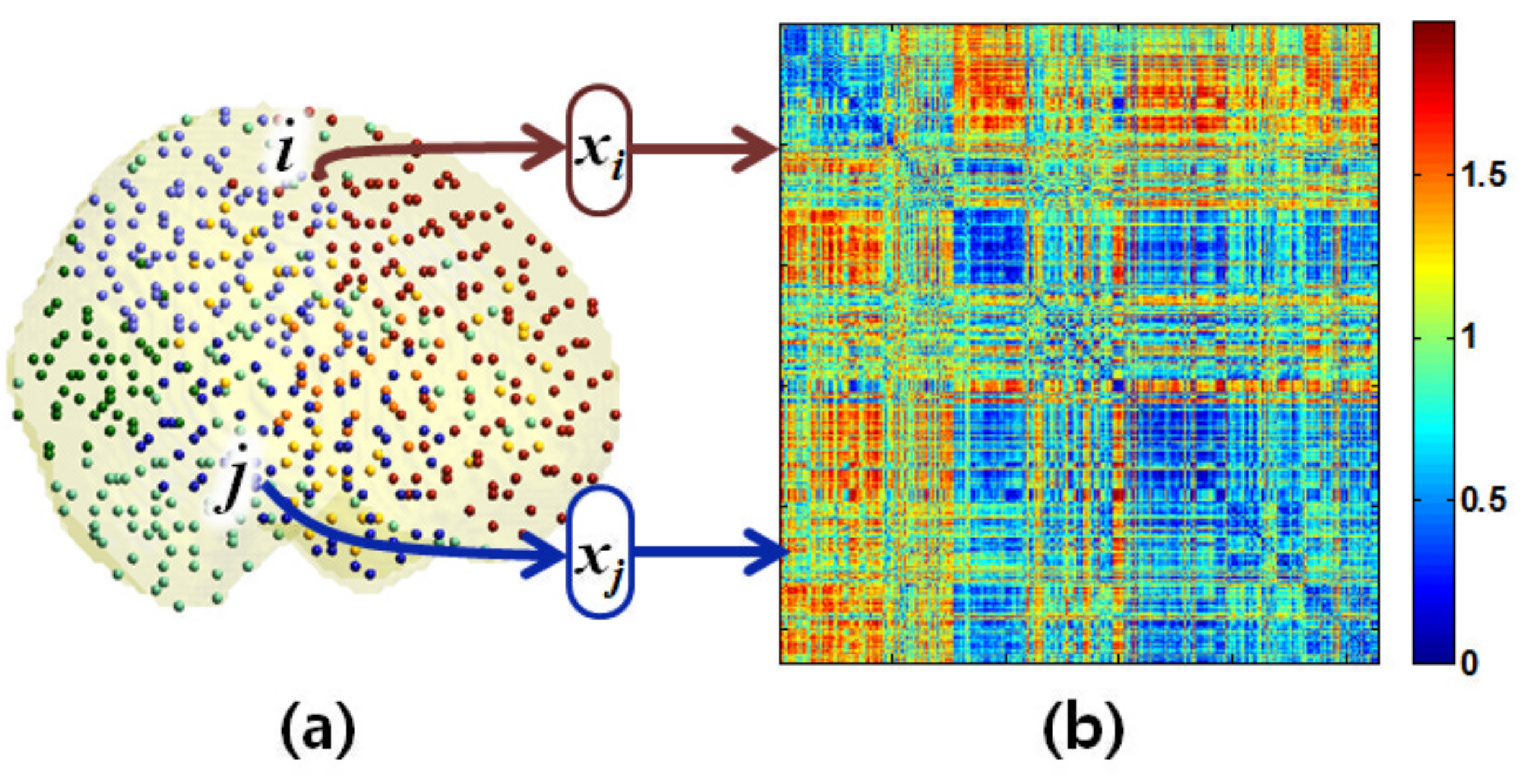}
\caption{Functional brain network in the metric space. (a) The network consists of the node set $V=\{1, \cdots, p \}$.
(b) At node $i$, we have vector measurement ${\bf x}_i$ corresponding to measurements from multiple FDG-PET images \cite{lee.2011.TMI}. The edge weights $\rho=(\rho_{ij})$ are often given by some bivariate function $f$, i.e., $\rho_{ij} = f({\bf x}_i, {\bf x}_j)$, which measures the strength of connectivity between brain regions.}
\label{fig:network_construction}
\vspace{-2cm}
\end{wrapfigure}

\begin{example}
\label{ex:corr}
Suppose we have measurement vector ${\bf x}_i  = (x_{1i}, \cdots, x_{ni})^{\top} \in \mathbb{R}^n$ on the node $i$ (Fig. \ref{fig:network_construction}). The weights $\rho=(\rho_{ij})$ between nodes is often given by some bivariate function $f$
on ${\bf x}_i$ and ${\bf x}_j$, i.e., $\rho_{ij} = f({\bf x}_i, {\bf x}_j)$. Denote the correlation between ${\bf x}_i$ and ${\bf x}_j$ as $\mbox{corr}({\bf x}_i, {\bf x}_j)$. If the weights $\rho=(\rho_{ij})$ are given by $\rho_{ij} = \sqrt{1-\mbox{corr}({\bf x}_i, {\bf x}_j)}$,  $\mathcal{X}=(V, \rho)$ forms a metric space.
\end{example}

\begin{example}
Suppose we center and rescale the measurement ${\bf x}_i$ such that $\| {\bf x}_i \|^2 = {\bf x}_i^{\top} {\bf x}_i =1$. Denote the full data matrix as ${\bf X}_{n \times p} = ({\bf x}_1, \cdots, {\bf x}_p)$. The edge weight of network  is then often given by some function of ${\bf X}$, i.e., $\rho = f({\bf X})$. For instance, the correlation $\rho_{ij}$ between the nodes $i$ and $j$ is given by ${\bf x}_i^{\top} {\bf x}_i$ and $\rho = {\bf X}^{\top} {\bf X}$ is the correlation matrix. The correlation is invariant under centering and scaling. Such operations are often done in practice to speed up the computation in sparse learning \cite{lee.2011.TMI,chung.2013.MICCAI,chung.2015.TMI}.  

Under centering and scaling, $\rho_{ij} = 1 - {\bf x}_i^{\top} {\bf x}_j$ {\em does not} satisfy the the triangle inequality. 
This is easily seen from the counter example
\bqn {\bf x}_i =(0, \frac{1}{\sqrt{2}},  -\frac{1}{\sqrt{2}})^{\top}, {\bf x}_j =(\frac{1}{\sqrt{2}},  0, -\frac{1}{\sqrt{2}})^{\top},  {\bf x}_k =(\frac{1}{\sqrt{6}}, \frac{1}{\sqrt{6}}, -\frac{2}{\sqrt{6}})^{\top} \label{eq:counter}\eqn
which satisfies 
$$ \| {\bf x}_i - {\bf x}_j \|^{2} > \| {\bf x}_i -  {\bf x}_k \|^{2} + \| {\bf x}_k -  {\bf x}_j \|^{2}$$
and results in $\rho_{ij} > \rho_{ik} + \rho_{jk}$.
$1 - \mbox{corr}$ is not a metric, but $\sqrt{1-\mbox{corr}}$ is a metric. 
\end{example}

Even though, the centered and rescaled correlation is no longer a metric in the Euclidean space, it may be a metric in some high dimensional non-Euclidean space. 

\begin{example}  
\label{ex:cos}
Under centering and scaling, we can allegorically show that $\rho({\bf x}_i, {\bf x}_j) = \cos^{-1}({\bf x}_i'{\bf x}_j)$ is a metric. This is equivalent to saying that 
${\bf x}_i'{\bf x}_j$ is a metric on the $n$-dimensional unit sphere. It obtains minimum $0$ when ${\bf x}_i'{\bf x}_j = 1$ and maximum $\pi$ when ${\bf x}_i'{\bf x}_j = -1$.
\end{example}

\section{Matrix norm as a network distance}

Matrix norm of the difference between networks is often used as measure of similarity between networks  \cite{banks.1994,zhu.2014}. Given two networks $\mathcal{X}^1=(V, \rho^1)$ and $\mathcal{X}^2=(V, \rho^2)$, the $L_l$-norm of network difference is given by
\bqn 
 D_l (\mathcal{X}^1,\mathcal{X}^2) = \parallel \rho^1 - \rho^2 \parallel_{l} =
            \Big(  \sum_{i,j}  \big| \rho^1_{ij} - \rho^2_{ij} \big|^{l}  \Big)^{1/l}. 
\label{eq:norm}
\eqn  
$D_{1}$ and $D_{2}$ are most often used and they are referred to as  $L_1$-distance and Euclidean distance ($L_2$). $L_2$-distance, often referred to as Frobenius norm distance, is probability the most often used matrix similarity measure \cite{zhu.2014} and can be rewritten as
$$D_2 (\mathcal{X}^1,\mathcal{X}^2) = \Big[ \mbox{tr } (\rho^1 -\rho^2)^2  \Big]^{1/2}.$$
If $\rho^1, \rho^2$ are adjacency matrices, $L_2$-distance is the Hamming metric often used in information theory \cite{banks.1994}. When $l=\infty,$ $L_{\infty}$-distance is written as 
\bqn 
 D_{\infty} (\mathcal{X}^1,\mathcal{X}^2) = \parallel \rho^1 - \rho^2 \parallel_{\infty} =
            \max_{\forall i,j}  \big| \rho^1_{ij} - \rho^2_{ij} \big|. 
\label{eq:inf_norm}
\eqn  
(\ref{eq:inf_norm}) is a special case of Gromov-Hausdorff (GH) distance and 
related to type-I error estimation under multiple comparisons \cite{chung.2013.CNA,lee.2017.HBM}.

The idea of using spherical geodesics in constructing the metric space in Example \ref{ex:cos} can be further extended to other abstract spaces. Consider a collection of positive definite symmetric matrices of size $p \times p$, which is denoted as $\mathcal{S}_p$. Many brain connectivity matrices including correlation and covariance metrics belong to $\mathcal{S}_p$ \cite{qiu.2015}. Given edge weights $ \rho^1, \rho^2 \in \mathcal{S}_p$, the shortest distance between $\rho^1$ and $\rho^2$ in this manifold is given by the log-Euclidean distance \cite{arsigny.2005}
$$d_{LE} (\rho^1, \rho^2) =  \left[ \mbox{tr} \left(  \log (\rho^1) - \log (\rho^2)   \right)^2 \right]^{1/2},$$
where $\log \cdot$ is the matrix logarithm. The log-Euclidean distance can be viewed as the generalized manifold version of Frobenius norm distance.

The element-wise differences  may not capture additional higher order similarity. For instance, there might be relations between a pair of columns or rows \cite{zhu.2014}. Further, $L_l$-distances usually surfer the problem of outliers. Since all the edges are assigned equal weights, few outlying extreme edge weights may severely affect the distance more than needed. Further, because they compare each edges separately, these distances simply ignore the underlying topological structures. It is necessary to define distances that are more topological in nature.

\begin{figure}[t]
\centering
\includegraphics[width=1\linewidth]{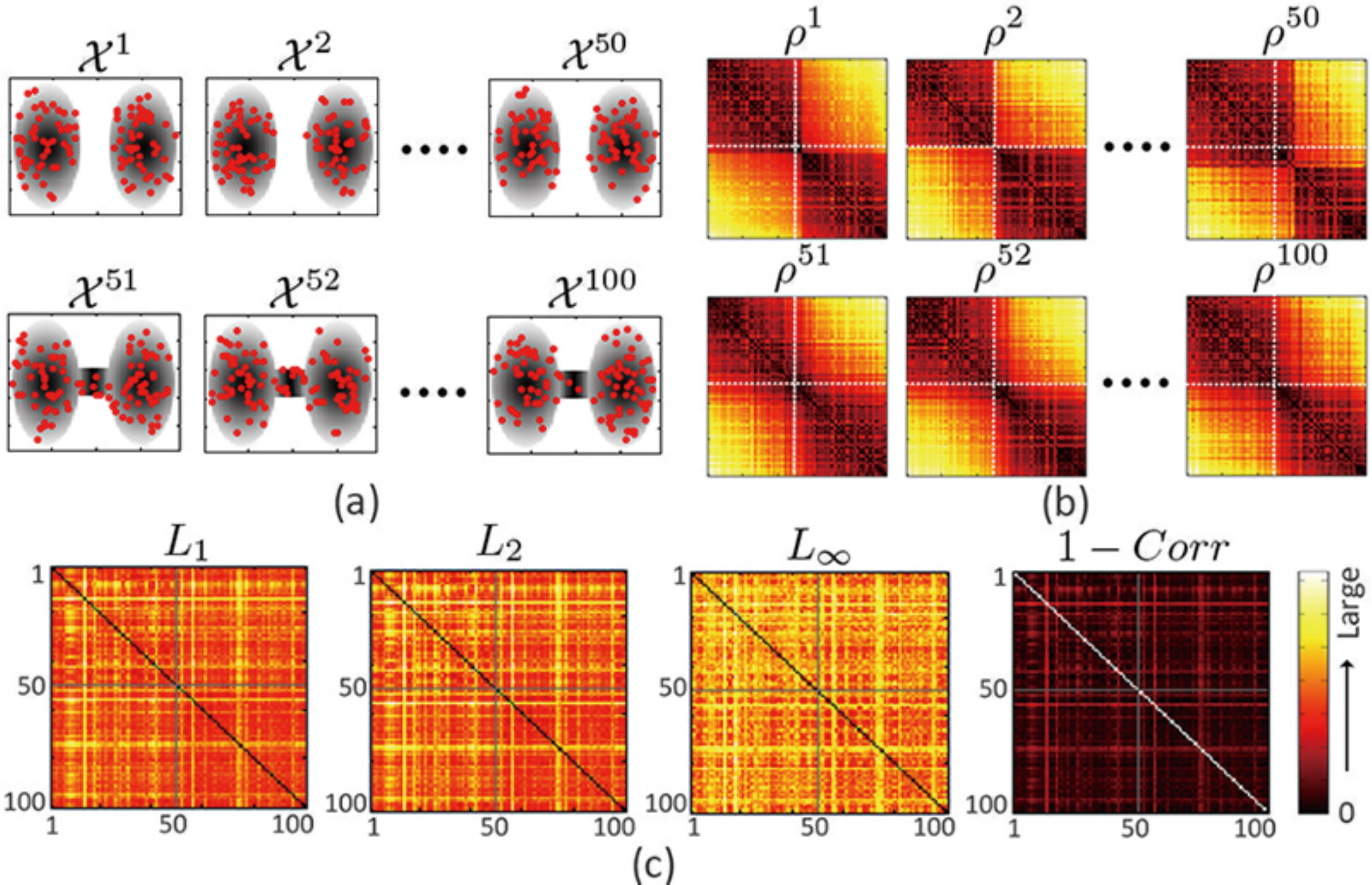}
\caption{Limitation of matrix norms as network distances. (a) $\mathcal{X}^1, \cdots, \mathcal{X}_{50}$ were simulated from the baseline network with two connected components. $\mathcal{X}_{51}, \cdots, \mathcal{X}_{100}$ were simulated from the baseline network with one connected component. Each network had 100 nodes. (b) Corresponding edge weights $\rho^{1},\cdots, \rho^{100}$ were given by the Euclidean distance between nodes. 
(c) Network distances based on $L_{1}$-, $L_{2}$-, $L_{\infty}$-distances, and 1 - Pearson correlation (from left to right).
Ward hierarchical clustering into two groups gives the clustering accuracy of 53 \%, 52 \%, 55 \%, and 50 \% respectively showing  poor performance. 
}
\label{fig:matrix_norm}
\end{figure}

\begin{example} 
\label{ex:matrix_norm} 
The insufficiency of $L_l$ network distances is illustrated in Fig. \ref{fig:matrix_norm}. Networks $\mathcal{X}^1, \cdots, \mathcal{X}_{50}$ were simulated from the baseline network with two connected components. Networks $\mathcal{X}_{51}, \cdots, \mathcal{X}_{100}$ were simulated from the baseline network with one connected component (Fig. \ref{fig:matrix_norm}-(a)). 
Each network has 100 nodes and their edge weights  $\rho^i$ are given by the Euclidean distance between nodes (b). 
The pairwise distance between 100 networks based on $L_{1}$-, $L_{2}$-, $L_{\infty}$-distances, and 1 - Pearson correlation are shown in (c). Ward hierarchical clustering \cite{lee.2011.MICCAI,ward.1963} into two groups gives the clustering accuracy of 53 \%, 52 \%, 55 \%, and 50 \%. 
The $L_l$ distances and correlations ignore the topological information of connectedness and result in extremely poor performance, which is no different from randomly flipping coins. This example demonstrates the need for topologically more aware distances in networks.  
\end{example}

\section{Network and graph filtrations}

To define topological distance in weighted networks in persistent homology,  it is necessary to construct a filtration. Let us threshold edge weights $\rho_{ij}$ and binarize a weighted network. 

\begin{definition} Given weighted network $\mathcal{X}=(V, \rho)$, the binary network $\mathcal{B}_{\epsilon} ( \mathcal{X}) =(V, \mathcal{B}_{\epsilon}(\rho))$ is a graph consisting of the node set $V$ and the edge weight $\mathcal{B}_{\epsilon}(\rho) = (\mathcal{B}_{\epsilon}(\rho_{ij}))$ given by 
\bqn \mathcal{B}_{\epsilon}(\rho_{ij}) =   \begin{cases}
1 &\; \mbox{  if } \rho_{ij} \leq \epsilon;\\
0 & \; \mbox{ otherwise}.
\end{cases}
\label{eq:case}
\eqn
\end{definition}

\begin{figure}[t]
\centering
\includegraphics[width=1\linewidth]{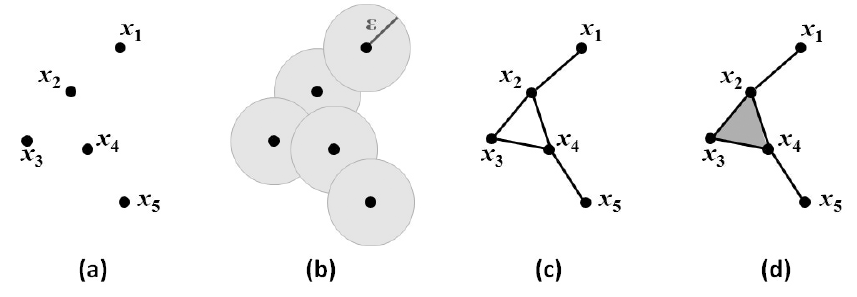}
\caption{The difference between a binary network and a Rips complex. (a) Point cloud data $X$. (b) The ball of radius $\epsilon$ centered at each point. (c) Binary network $\mathcal{B}_{\epsilon}(X)$ (d) Rips complex $\mathcal{R}_{\epsilon}(X)$. Unlike the binary network, the Rips complex has a filled-in triangle.}
\label{fig:Rips1}
\end{figure}

Note $\mathcal{B}_{\epsilon}(\rho)$ is the adjacency matrix of $\mathcal{B}_{\epsilon} ( \mathcal{X})$. The binary network $\mathcal{B}_{\epsilon}(\mathcal{X})$ is a simplicial complex consisting of $0$-simplices (nodes) and $1$-simplices (edges)  \cite{erickson.2009.notes,ghrist.2008.bams}. 
In the metric space $\mathcal{X}=(V, \rho)$, the Rips complex $\mathcal{R}_{\epsilon}(X)$ is a simplicial complex whose $(p-1)$-simplices correspond to unordered $p$-tuples of points that satisfy $\rho_{ij} \leq \epsilon$ in a pairwise fashion \cite{ghrist.2008.bams}.
While the binary network $\mathcal{B}_{\epsilon}(\mathcal{X})$ has at most 1-simplices, the Rips complex can have at most $(p-1)$-simplices (Fig. \ref{fig:Rips1}). Thus, we have $\mathcal{B}_{\epsilon}(\mathcal{X}) \subset \mathcal{R}_{\epsilon}(\mathcal{X})$. The Rips complex has the property that
$$\mathcal{R}_{\epsilon_0}(\mathcal{X}) \subset \mathcal{R}_{\epsilon_1}(\mathcal{X}) \subset \mathcal{R}_{\epsilon_2}(\mathcal{X}) \subset \cdots $$
for $0=\epsilon_{0} \le \epsilon_{1} \le \epsilon_{2} \le \cdots.$
When $\epsilon=0$, the Rips complex is simply the node set $V$. By increasing the filtration value $\epsilon$, we are connecting more nodes so the size of the edge set increases.
Such the nested sequence of the Rips complexes is called a Rips filtration, the main object of interest in the persistent homology \cite{edelsbrunner.2009}.

Since a binary network is a special case of the Rips complex, it inherits all the topological properties of the Rips complex including filtration $$\mathcal{B}_{\epsilon_0} (\mathcal{X}) \subset \mathcal{B}_{\epsilon_1} (\mathcal{X}) \subset \mathcal{B}_{\epsilon_2} (\mathcal{X}) \subset \cdots  $$
for
$0=\epsilon_{0} \le \epsilon_{1} \le \epsilon_{2} \cdots.$
The sequence of such nested multiscale graph structure  is called the {\em graph filtration} \cite{lee.2011.MICCAI,lee.2012.TMI}.

\begin{figure}[t]
\centering
\includegraphics[width=1\linewidth]{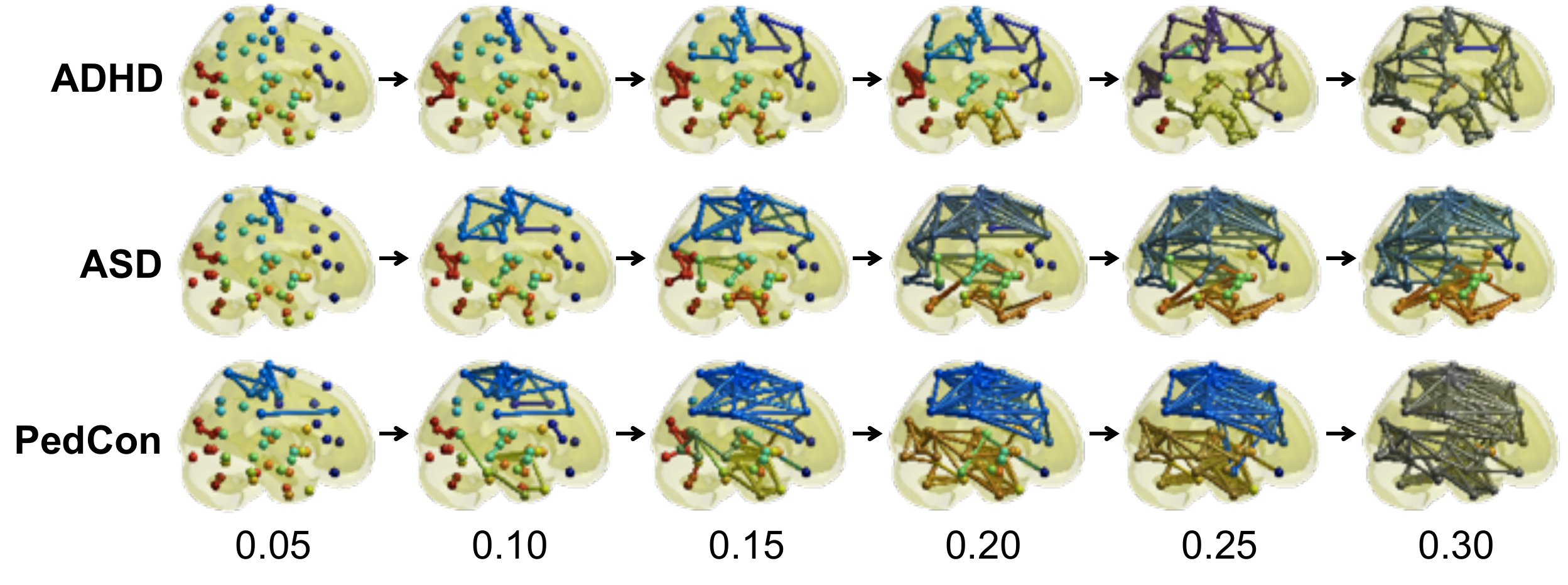}
\caption{Graph filtrations of the inter-subject correlation in FDG-PET measurements showing group differences between attention-deficit hyperactivity disorder (ADHD), autism spectrum disorder (ASD) children and pediatric controls (PedCon) \cite{lee.2012.TMI}. The filtration value is $1 -$ correlation. 
PedCon clerly shows much dense brain connectivity over the filtration.}
\label{fig:graphfiltration1}
\end{figure}

\begin{example}
Fig. \ref{fig:graphfiltration1} shows an example of a graph filtration in fluorodeoxyglucose (FDG) positron emission tomography (PET) study of three different populations \cite{lee.2012.TMI}. Graph filtrations of the inter-subject correlation in FDG-PET measurements showing group differences between attention-deficit hyperactivity disorder (ADHD), autism spectrum disorder (ASD) children and pediatric controls (PedCon). The  $1 -$ Pearson correlation is used as filtration values. PedCon clearly shows much dense brain connectivity over the filtration.
\end{example}

If we order the edge weights in an increasing order, we have the sorted edge weights:
$$\min_{i,j} \rho_{ij} = \rho_{(1)} \leq \rho_{(2)} \leq  \cdots \leq \rho_{(q)} = \max_{i,j} \rho_{ij},$$
where $q \leq (p^2-p)/2$, the maximum possible number of edge weights. The subscript $_{( \;)}$ denotes the order statistic.

Due to multiplicity of edge weights, the filtration 
$$\mathcal{B}_0 (\mathcal{X}) \subset \mathcal{B}_{\rho_{(1)}} (\mathcal{X}) \subset  \cdots  \subset  \mathcal{B}_{\rho_{(q)}} (\mathcal{X})$$
may {\em not} be unique in a sense we can easily have
$ \mathcal{B}_{\rho_{(i)}} (\mathcal{X})= \mathcal{B}_{\rho_{(i+1)}} (\mathcal{X})$ for some $\rho_{(i)} = \rho_{(i+1)}$. To avoid this redundancy, we sort and remove any multiplicative weights and obtain the set of unique positive weights 
$$\rho_{u(1)} < \rho_{u(2)} <  \cdots <  \rho_{u(r)}.$$
The graph filtration on this unique edge set will be unique:

\begin{theorem} 
\label{theorem:maximal}
For graph $X=(V, \rho)$ with $r$ distinct edge weights $0 < \rho_{u(1)} < \rho_{u(2)} < \cdots < \rho_{u(r)}$ can  be uniquely decomposed as the graph filtration 
\bqn \mathcal{B}_{0} (\mathcal{X}) \subset \mathcal{B}_{\rho_{u(1)}} (\mathcal{X}) \subset \cdots \subset  \mathcal{B}_{\rho_{u(r)}} (\mathcal{X}).\label{eq:maximal}\eqn
\end{theorem}

The proof of uniqueness is given in \cite{chung.2015.TMI}. For any $\epsilon \geq 0$,  
$\mathcal{B}_{\epsilon} (\mathcal{X})$ belongs to one of the binary network in the filtration (\ref{eq:maximal}). In this sense, the representation (\ref{eq:maximal}) is unique. The edge weights $\rho_{ij}$ can be recovered from the filtration and the filtration values. Let $a_{ij}^k = \rho_{u(k)}$ if nodes $i$ and $j$ are connected in binary network $\mathcal{B}_{\rho_{u(k)}}(\mathcal{X})$ and $a_{ij}^k = \infty$ otherwise. Then 
$$\rho_{ij} = \min ( a_{ij}^{1},a_{ij}^{2}, \cdots, a_{ij}^{r}).$$
The finiteness and uniqueness of the filtration over weighted graphs are intuitively clear by themselves and are already applied in software packages such as javaPlex \cite{adams.2014}.

\section{Bottleneck distance}

The network topology can be represented by the birth and death of homology groups in persistent homology. The zeroth and first homology groups are a set of connected components and holes, respectively. 
During a filtration, homologies in homology group appear and disappear. 
If a homology appears at the threshold $\xi$ and disappears at $\tau,$ it can be encoded into a point, $(\xi,\tau)  \; (0 \le \xi \le \tau < \infty)$ in $\mathbb{R}^2$. If $m$ number of homologies appear during the filtration of a network $\mathcal{X}=(V,\rho)$, the homology group can be represented by a point set $$\mathcal{P} (\mathcal{X}) = \left\{ (\xi_{1},\tau_{1}), \dots, (\xi_{m},\tau_{m}) \right\}.$$ This scatter plot is called the persistence diagram (PD) \cite{cohensteiner.2007}. Note that the Betti number is the cardinality of homology group at the fixed threshold, while $m$ is the number of homologies that 
appear during the filtration.

Given two networks $\mathcal{X}^1=(V,\rho^1)$ and $\mathcal{X}^2=(V,\rho^2),$ we construct the corresponding graph filtrations. Subsequently,PDs $$\mathcal{P} (\mathcal{X}^1) = \left\{ (\xi_{1}^{1},\tau_{1}^{1}), \cdots, (\xi_{m}^{1},\tau_{m}^{1}) \right\}$$ and $$\mathcal{P} (\mathcal{X}^2) = \left\{ (\xi_{1}^{2},\tau_{1}^{2}), \cdots, (\xi_{n}^{2},\tau_{n}^{2}) \right\}$$ are obtained through the filtration. 

The bottleneck distance between networks is defined as the bottleneck distance of the corresponding PDs: 
\bqn 
D_{B} (\mathcal{P} (\mathcal{X}^1),\mathcal{P} (\mathcal{X}^2)) = 
\inf_{\gamma} \sup_{1 \leq i \leq m} \parallel t_{i}^{1} - \gamma(t_{i}^{1}) \parallel_{\infty},
\label{eq:D_B}
\eqn 
where $t_{i}^{1} = (\xi_{i}^{1},\tau_{i}^{1}) \in \mathcal{P} (\mathcal{X}^1)$, $t_{j}^{2} = (\xi_{j}^{2},\tau_{j}^{2}) \in \mathcal{P} (\mathcal{X}^2)$  and $\gamma$ is a bijection from $\mathcal{P} (\mathcal{X}^1)$ to $\mathcal{P} (\mathcal{X}^2)$. If  $t_{j}^{2} = \gamma(t_{i}^{1})$,
the infinity norm is defined as $\parallel t_{i}^{1} - \gamma(t_{i}^{1}) \parallel_{\infty} = \max ( | \xi_{i}^{1}-\xi_{j}^{2}|,| \tau_{i}^{1}-\tau_{j}^{2}|).$ 
Note (\ref{eq:D_B}) assumes $m=n$ such that the bijection $\gamma$ exists. If $m \neq n$, there is no one-to-one correspondence between two PDs. Then, auxiliary points  $$(\frac{\xi_{1}^{1}+\tau_{1}^{1}}{2},\frac{\xi_{1}^{1}+\tau_{1}^{1}}{2}), \cdots, (\frac{\xi_{m}^{1}+\tau_{m}^{1}}{2},\frac{\xi_{m}^{1}+\tau_{m}^{1}}{2})$$ and $$(\frac{\xi_{1}^{2}+\tau_{1}^{2}}{2},\frac{\xi_{1}^{2}+\tau_{1}^{2}}{2}), \cdots, (\frac{\xi_{n}^{2}+\tau_{n}^{2}}{2},\frac{\xi_{n}^{2}+\tau_{n}^{2}}{2})$$ that are orthogonal projections to  the diagonal line $\xi = \tau$ of points in $\mathcal{P} (\mathcal{X}^1)$ and $\mathcal{P} (\mathcal{X}^2)$ are added to  $\mathcal{P} (\mathcal{X}^2)$ to $\mathcal{P} (\mathcal{X}^1)$ respectively to make the identical number of points in PDs. 
The bijection  $\gamma$ is determined by the bipartite graph matching algorithm \cite{cohensteiner.2007}. 
Since $\gamma$ is a one-to-one correspondence with the minimum sum of distances, points that are close to the diagonal line, i.e., those with small durations, are predominantly mapped to auxiliary points. Thus, the auxiliary points are usually ignored during the estimation of bottleneck distance. This is consistent with the premise of persistent homology that homology with a longer duration is a signal of the topological space, but homology with a shorter duration is noise.

\begin{figure}[t]
\centering
\includegraphics[width=1\linewidth]{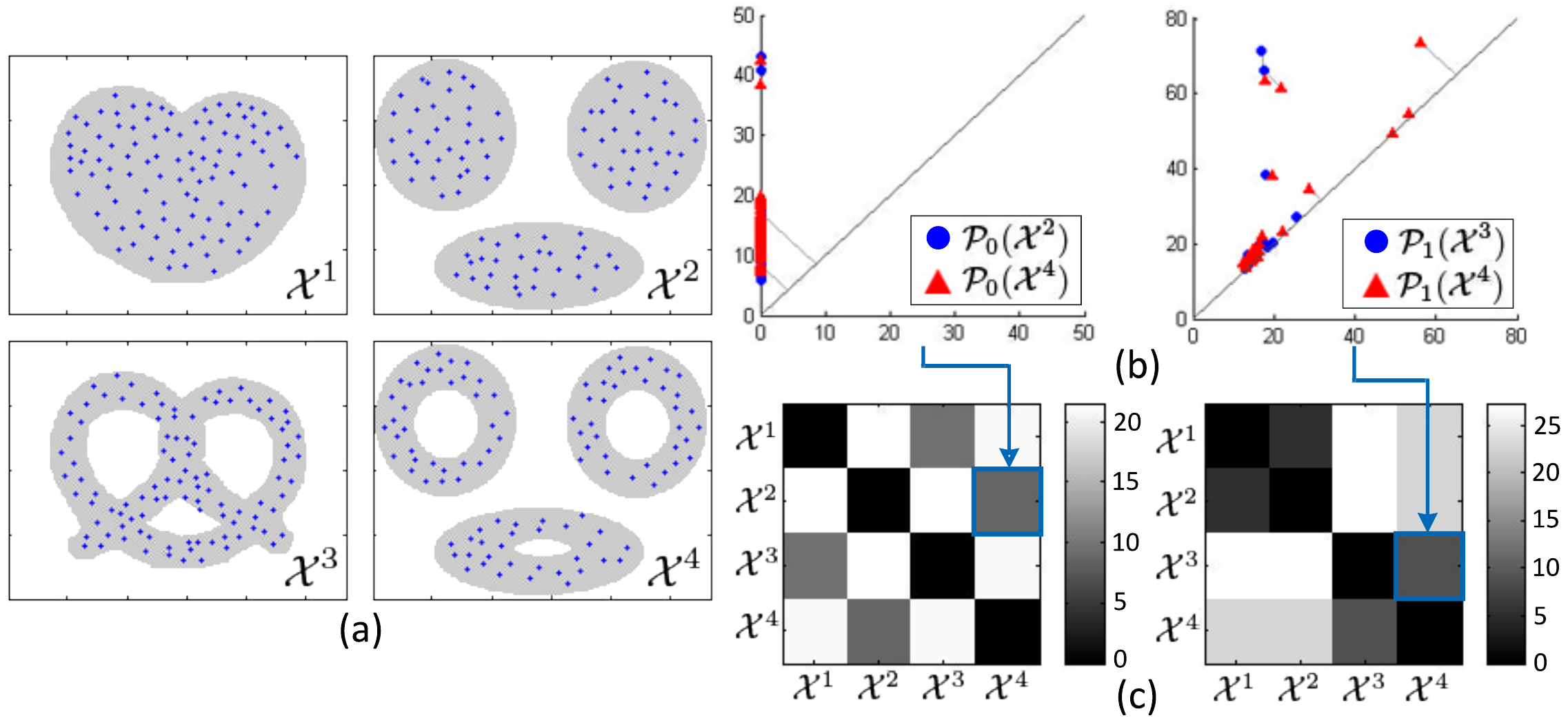}
\caption{(a) Examples of topological spaces, (b) the PDs of $\mathcal{X}^2$ and $\mathcal{X}^4$ of connected components, denoted as $\mathcal{P}_{0} (\mathcal{X}^2)$ and $\mathcal{P}_{0} (\mathcal{X}^4)$ (left) and one of $\mathcal{X}^3$ and $\mathcal{X}^4$ of holes, denoted as $\mathcal{P}_{1} (\mathcal{X}^3)$ and $\mathcal{P}_{1} (\mathcal{X}^4)$ (right), (c) the bottleneck distances of connected components (left) and holes (right).}
\label{fig:example_bottleneck}
\end{figure}

\begin{example}
\label{ex:bottleneck}
Examples of bottleneck distance are shown in Fig. \ref{fig:example_bottleneck}. Given four topological spaces $\mathcal{X}^1, \mathcal{X}^2, \mathcal{X}^3, $ and $\mathcal{X}^4,$ the PDs of connected components and holes are shown in (b). 
The bottleneck distance can easily distinguish $\mathcal{X}^1, \mathcal{X}^3$ (one connected component) from $ \mathcal{X}^2, \mathcal{X}^4$ (three connected components)
as shown in (c)-left. 
When it is applied to the PDs of holes, it can easily distinguish $\mathcal{X}^1, \mathcal{X}^2$ (no hole) from $\mathcal{X}^3, \mathcal{X}^4$ (three holes) as shown in (c)-right.
\end{example}

\begin{example}
When the bottleneck distance is applied to 100 simulated networks in Example \ref{ex:matrix_norm}, the networks are clustered according to two baseline topological structures with  100 \% accuracy. The superior almost perfect performance is due to the fact that  the bottleneck distance use the topological information of connectedness represented in PD.
\end{example}

\section{Kernel distance}
\label{sec:kernel_distance}

A multi-scale kernel method for PDs based on state space theory has been proposed in \cite{pachauri.2011,reininghaus.2015}. 
It embeds the topological information in PD into a feature space by kernel smoothing of PD $\mathcal{P} (\mathcal{X}) = \{t_1, \cdots, t_m \}$:  
\bq
\Phi_{\sigma} (\mathcal{P} (\mathcal{X})) = \frac{1}{4 \pi \sigma} \sum_{i=1}^m \exp{\left( -\frac{\parallel t-t_{i} \parallel_{2}^{2}}{4 \sigma} \right)} - \exp{ \left( -\frac{\parallel t-\bar{t_{i}} \parallel_{2}^{2}}{4 \sigma} \right)}, 
\eq
where 
$\bar{t_{i}} = (\tau_{i}, \xi_{i})$ is the mirror reflection of $t_{i} = (\xi_{i},\tau_{i})$ with respect to the diagonal line.
Then the inner product between two PDs is given by kernel 
\bqn 
k_{\sigma} (\mathcal{P} (\mathcal{X}^1),\mathcal{P} (\mathcal{X}^2)) &=& \left<  \mathcal{P} (\mathcal{X}^1),\mathcal{P} (\mathcal{X}^2) \right> = \int_{\xi \ge \tau}\Phi_{\sigma} (\mathcal{P} (\mathcal{X}^1)) \Phi_{\sigma} (\mathcal{P} (\mathcal{X}^2)) dt  \nonumber \\ 
&=& \frac{1}{8 \pi \sigma} \sum_{i,j} \exp{\left( -\frac{\parallel t_{i}^{1} - t_{j}^{2} \parallel_{2}^{2}}{8 \sigma} \right)} - \exp{ \left(-\frac{\parallel t_{i}^{1} - \bar{t_{j}^{2}} \parallel_{2}^{2}}{8 \sigma} \right)}. \nonumber 
\eqn  
$k_{\sigma}$ is called the persistence state-space (PSS) kernel. 
Kernel distance $D_{K}$ is then defined as a distance between PDs in a feature space \cite{scholkopf.2001}:
\bq
D_{K}  (\mathcal{P} (\mathcal{X}^1),\mathcal{P} (\mathcal{X}^2)) &=& \parallel \Phi_{\sigma} (\mathcal{P} (\mathcal{X}^1)) - \Phi_{\sigma} (\mathcal{P} (\mathcal{X}^2)) \parallel \\
&=& \left<  \Phi_{\sigma} (\mathcal{P} (\mathcal{X}^1)) - \Phi_{\sigma} (\mathcal{P} (\mathcal{X}^2),\Phi_{\sigma} (\mathcal{P} (\mathcal{X}^1)) - \Phi_{\sigma} (\mathcal{P} (\mathcal{X}^2) \right>^{1/2} \\ 
&=& \big[   k_{\sigma} (\mathcal{P} (\mathcal{X}^1), \mathcal{P} (\mathcal{X}^1)) 
+ k_{\sigma} (\mathcal{P} (\mathcal{X}^2),\mathcal{P} (\mathcal{X}^2)) \\
&& - 2 k_{\sigma} (\mathcal{P} (\mathcal{X}^1),\mathcal{P} (\mathcal{X}^2)) \big]^{1/2}. 
\eq
The performance of kernel distance depends on the parameter $\sigma.$ Thus, $\sigma$ should be chosen carefully.  

\begin{figure}[t]
\centering
\includegraphics[width=1\linewidth]{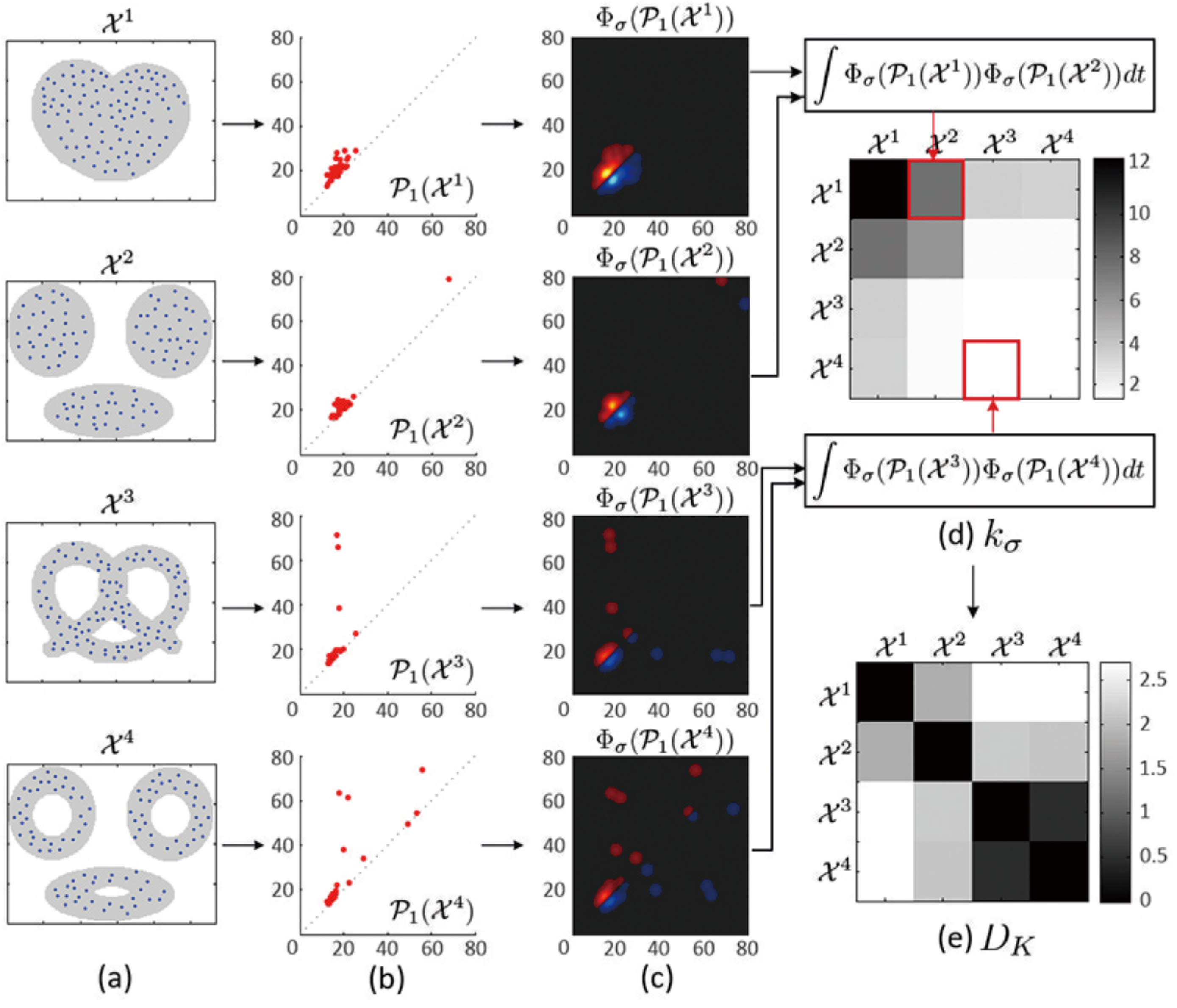}
\caption{(a) Same topological spaces given in Fig. \ref{fig:example_bottleneck} are used. (b) PDs of holes $\mathcal{P}_{1} (\mathcal{X}^i)$ are computed, (c) kernel smoothing of PDs with the additional mirror image with respect to the diagonal line, denoted as $\Phi_{\sigma}(\mathcal{P}_{1} (\mathcal{X}^i))$, (d) persistence state-space kernel $k_{\sigma}$, (e) kernel distance $D_{K}$.}
\label{fig:example_kernel}
\end{figure}

\begin{example}The same topological spaces given in Example \ref{ex:bottleneck} are used. The PDs of holes are shown in Fig. \ref{fig:example_kernel}-(b). After kernel smoothing of PDs (c), the kernel distances $D_K$ are estimated (d, e).    
As in the bottleneck distances in Fig. \ref{fig:example_bottleneck}-(c), the kernel distance in (e) can distinguish $\mathcal{X}^1, \mathcal{X}^2$ (no hole) from $ \mathcal{X}^3, \mathcal{X}^4$ (three holes). 
\end{example}

The bottleneck distance does not  take into account the difference in the number of connected components or holes since it compares two PDs through a one-to-one correspondence. However, the kernel distance estimates the difference of all pairs of points in two PDs. Moreover, the bottleneck distance tends to ignore points that are close to the diagonal line in PD by mapping to auxiliary points although such points may sometimes indicate the underlying topological structure that discriminates topological spaces.

\section{Gromov-Hausdorff distance} 
\label{sec:gh_distance}

\begin{figure}[t]
\centering
\includegraphics[width=1\linewidth]{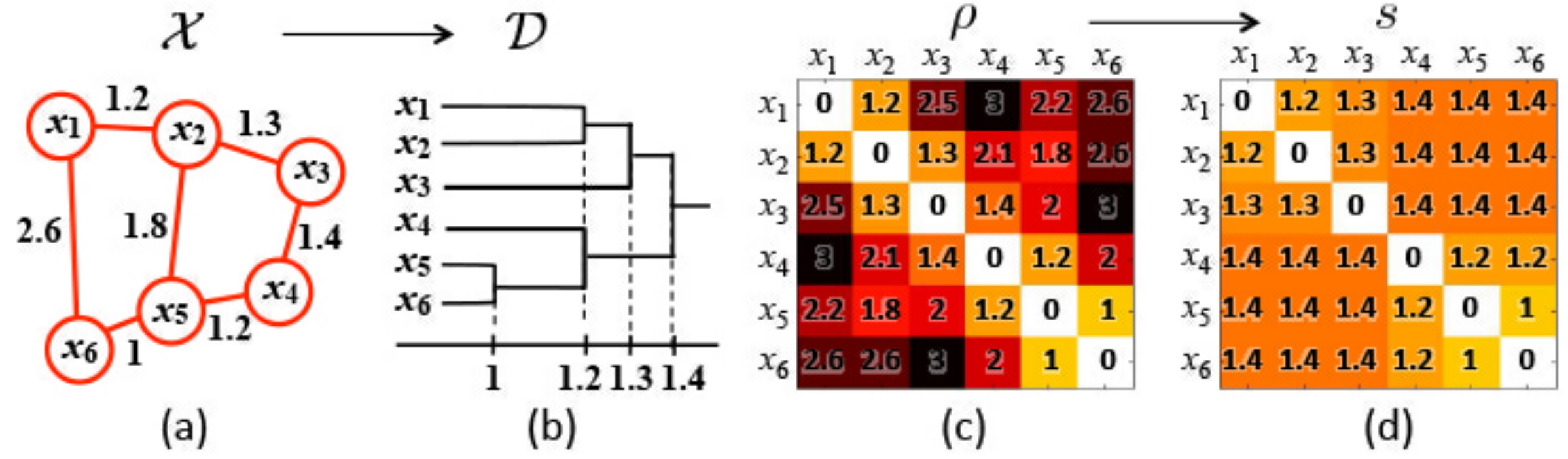}
\caption{(a) An example of a network, (b) its dendrogram, (c) the distance matrix based on Euclidean distance, (d) the single linkage matrix (SLM).}
\label{fig:example_GHdistance}
\end{figure}

The change of connected components during the filtration has an agglomerative hierarchical clustering structure based on single linkage \cite{carlsson.2008,carlsson.2010.jmlr,lee.2012.TMI}. 
Each node starts as a single connected component, and pairs of connected components then are merged if the distance between the closest nodes in two disjoint connected components is smaller than the filtration value. The distance $s_{ij}$ between the closest nodes in the two disjoint connected components ${\bf R}_1$ and ${\bf R}_2$ is called the single linkage distance (SLD), which is defined as 
$$ s_{ij} = \min_{l \in {\bf R}_{1}, k \in {\bf R}_{2}} \rho_{lk}.$$ 
Thus, every edge connecting a node in ${\bf R}_1$ to a node in ${\bf R}_2$ has the same SLD.
SLD is then used to construct the single linkage matrix (SLM) $S = (s_{ij})$ (Fig. \ref{fig:example_GHdistance}).  
SLM shows how connected components are merged locally and can be used in constructing a dendrogram. SLM is a {\em ultrametric}
 which is a metric space satisfying the stronger triangle inequality $s_{ij} \le \max (s_{ik},s_{kj})$ \cite{carlsson.2010.jmlr}.
Thus the dendrogram can be represented as a ultrametric space $\mathcal{D} = (V,S),$ which is again a metric space. 

\begin{figure}[t]
\begin{center}
\includegraphics[width=0.7\linewidth]{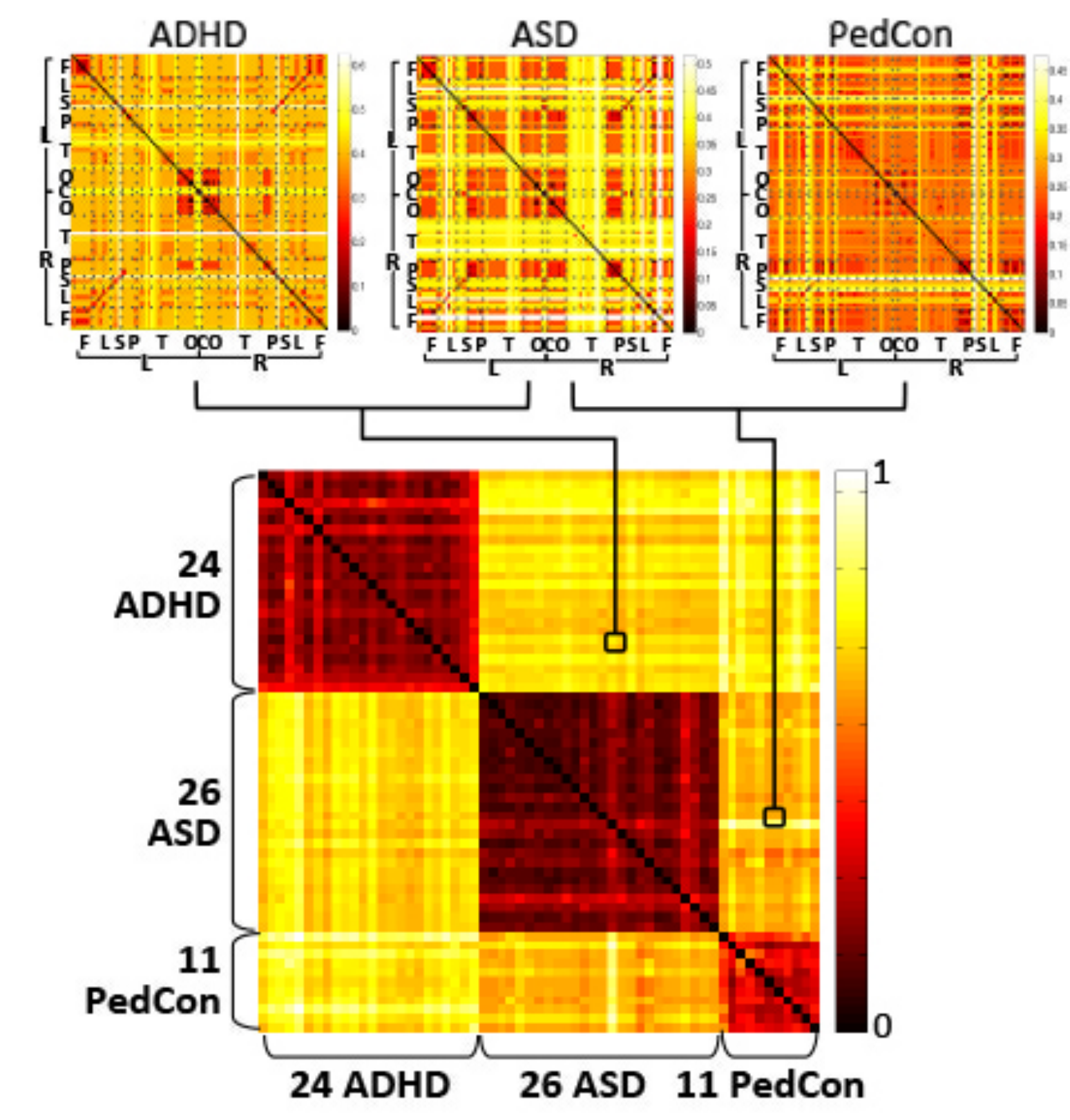}
\caption{SLMs of ADHD, ASD and PedCon groups of FDG-PET brain networks (top).  Since each group gives a single SLM, the jackknife resampling was used to replicate SLMs. GH distance matrix shows the pairwise distance between networks, which is half the difference between SLMs. Three diagonal block matrices shows distinct clusters between the three groups.}
\label{fig:GH_distance}
\end{center}
\end{figure}

GH distance between networks can be defined as GH distance between corresponding dendrograms $\mathcal{D}^{1}=(V,S^{1})$ and $\mathcal{D}^{2}=(V,S^{2})$ with SML $S^1 = (s^1_{ij})$ and $S^2 = (s^2_{ij})$:
\bqn D_{GH} (\mathcal{D}^1,\mathcal{D}^2) = \frac{1}{2} \max_{\forall i,j}  | s^1_{ij} - s^2_{ij} |. \label{eq:D_GH} \eqn
Although it is possible to define GH distance directly from the metric spaces $\mathcal{X}^{1}=(V,\rho^{1})$ and $\mathcal{X}^{2}=(V,\rho^{2})$, it turns out that it is identical to $L_{\infty}$ distance between networks in (\ref{eq:inf_norm}) thus ignoring the topological structures of the network. This is the main reason the GH distance is defined through the dendrograms \cite{lee.2011.MICCAI,lee.2012.TMI}.

\begin{example}
When GH distance is applied to 100 simulated networks in Example \ref{ex:matrix_norm}, the networks are clustered according to two baseline topological structures with  100 \% accuracy. The superior almost perfect performance is due to the fact that  GH distance use the topological information of connectedness represented in the corresponding dendrograms.
\end{example}

\begin{example} GH distance (\ref{eq:D_GH}) is used in discriminating brain networks in \cite{lee.2011.MICCAI,lee.2012.TMI}. FDG-PET data consists of 24 ADHD, 26 ASD and 11 pediatric controls (PedCon). Each group gives a single SLM (Fig. \ref{fig:GH_distance}). The jackknife resampling is applied in replicating 24, 26, and 11 SLMs from ADHD, ASD, and PedCon respectively. GH distance is used to compute the distance between SLMs in a pairwise fashion. Based on the GH distance matrix, SLMs were clustered into 3 groups using Ward hierarchical clustering. The clustering accuracy was $100$ \%.
\end{example}

While the bottleneck distance measures the difference between networks by encoding their topological information into the PDs, GH distance measures the difference between networks by embedding the network into the ultrametric space that represents hierarchical clustering structure of network \cite{carlsson.2010.jmlr}. GH distance can be viewed as measuring distance between corresponding dendrograms of the networks. GH distance has been statistically quantified using resampling techniques such as jackknife or permutation tests \cite{lee.2011.MICCAI,lee.2012.TMI,lee.2017.HBM}.

\section{KS-test like distance}


The graph filtration can be quantified using monotonic function $f$ satisfying
\bqn  f \circ \mathcal{B}_0 (\mathcal{X}) \geq f \circ \mathcal{B}_{\rho_{(1)}} (\mathcal{X}) \geq  \cdots  \geq f \circ \mathcal{B}_{\rho_{(q)}} (\mathcal{X}).\label{eq:B} \eqn
The number of connected components, the zeroth Betti number $\beta_0$, satisfies the monotonicity property (\ref{eq:B}). The size of the largest cluster satisfies a similar but opposite relation of monotonic increase. Many graph theoretic features such as node degree will be also monotonic. However, any other higher order Betti numbers may not be monotonic.  The monotonic feature vectors (\ref{eq:B}) are used in quantifying the brain networks \cite{chung.2013.MICCAI,chung.2015.TMI,lee.2011.ISBI}. In \cite{lee.2011.ISBI}, the slope of the linear regression line is fitted in  scatter points $(\rho_{(j)},  f \circ \mathcal{B}_{\rho_{(j)}} (\mathcal{X}))_{1 \leq j \leq q}$ and used to differentiate different clinical populations.  In  \cite{chung.2015.TMI}, integral $\int_{0}^{\infty} f \circ \mathcal{B}_{\epsilon} (\mathcal{X}) \;d\epsilon$ is used as the summary statistic instead of the vector (\ref{eq:B}).

Given two networks $\mathcal{X}^1=(V, \rho^1)$ and $\mathcal{X}^2=(V, \rho^2)$, 
Kolmogorove-Smirnov (KS) test like distance between $\mathcal{X}^1$ and $\mathcal{X}^2$ is given by
$$D_{KS}(\mathcal{X}^1, \mathcal{X}^2) = \sup_{\epsilon \geq 0} \big| f\circ \mathcal{B}_{\epsilon} (\mathcal{X}^1) - f \circ \mathcal{B}_{\epsilon}(\mathcal{X}^2) \big|,$$
where $f$ is a monotonic topological feature \cite{chung.2013.MICCAI,chung.2015.TMI,lee.2017.HBM}. $D_{KS}$ is motivated by the two-sample Kolmogorove-Smirnov (KS) test \cite{bohm.2010,chung.2013.MICCAI,gibbons.1992}, which is a nonparametric test for determining the equivalence of two cumulative distribution functions (CDF). The curve $(\epsilon,  f \circ \mathcal{B}_{\epsilon} (\mathcal{X}))_{\epsilon \geq 0}$ looks exactly like a CDF if it is properly normalized (Fig. \ref{fig:KS-example}). The number of connected components $\beta_0$ and the size of the largest connected component can be used for $f$. However, $\beta_1$ cannot be used since it is not monotonic. 

The distance $D_{KS}$ can be discretely approximated using the finite number of filtrations:
$$D_q = \sup_{1 \leq j \leq q} \big| f\circ \mathcal{B}_{\epsilon_j} (\mathcal{X}^1) - f \circ \mathcal{B}_{\epsilon_j}(\mathcal{X}^2) \big|.$$
If we choose enough number of $q$ such that $\epsilon_j$ are all the sorted edge weights, then $D_{KS}(\mathcal{X}^1,\mathcal{X}^2) = D_q$ due to Theorem \ref{theorem:maximal}. This is possible since there are only up to $p(p-1)/2$ number of unique edges in a graph with $p$ nodes and $ f\circ B$ increase discretely. In practice,  $\epsilon_j$ may be chosen uniformly or a divide-and-conquer strategy can be used to do adaptively grid the filtration values.

\begin{figure}[t]
\centering
\includegraphics[width=1\linewidth]{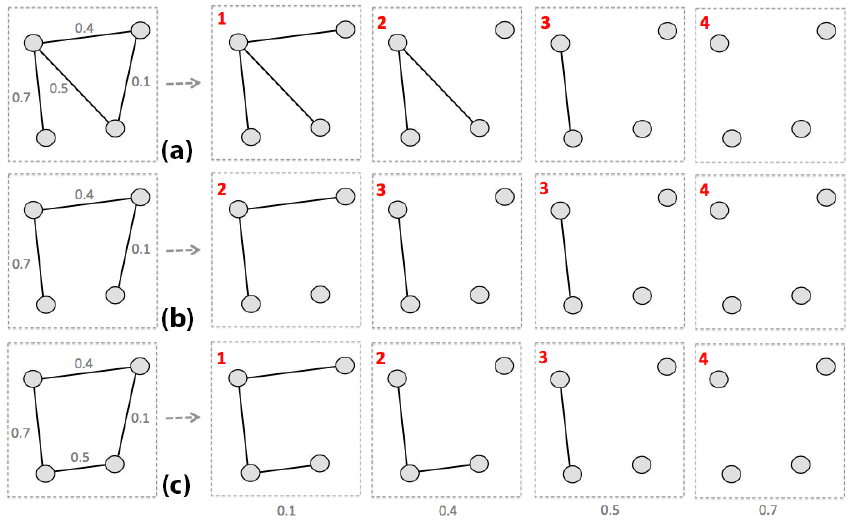}
\caption{The graph filtration. The zeroth Betti number $\beta_0$ (red numbers) is used to compute the network distance. The distance between graphs $(a)$ and $(b)$ is 1 and it is the maximum difference in $\beta_0$-values between the filtrations. The network distance between $(a)$ and $(c)$ is 0. Note $(a)$ and $(c)$ are topologically equivalent and there is no difference in $\beta_0$ and $\beta_1$.}
\label{fig:distsup}
\end{figure}

\begin{example} Fig. \ref{fig:distsup} illustrates graph filtrations and the KS-test like distance using $\beta_0$-values (red numbers) as $f$. The distance between graphs $(a)$ and $(b)$ is 1. $D_q$ easily capture the topological differences between $(a)$ and $(b)$. The distance between $(a)$ and $(c)$ is 0. There is no topological difference between $(a)$ and $(c)$. Both has a cycle and one connected component. On the other hand, GH distance \cite{lee.2011.MICCAI} is unable to distinguish between $(a)$ and $(b)$ and treat $(a)$ and $(c)$ as different networks. The proposed graph distance seems to be more topologically aware than the GH distance in when cycles are present.
 \end{example}

\section{Statistical inference on distances}

Given two networks $\mathcal{X}^1 = (V, \rho^1)$ and $\mathcal{X}^2 = (V, \rho^2)$, we are interested in testing the null hypothesis $H_0$ of the equivalence of two networks
$$H_0: \mathcal{X}^1 = \mathcal{X}^2.$$ 
This is probably the most often encountered hypothesis setting in brain imaging applications. This is still a challenging inference problem and it is not completely resolved in general cases.  We will focus on a much smaller but tractable problem of testing if two monotonic graph functions are equivalent:
\bq H_0':  f\circ \mathcal{B}_{\epsilon} (\mathcal{X}^1) &=& f \circ \mathcal{B}_{\epsilon}(\mathcal{X}^2) \; \mbox{ for all } \epsilon \geq 0\\
& vs. &\\
H_1':  f\circ \mathcal{B}_{\epsilon} (\mathcal{X}^1) &\neq& f \circ \mathcal{B}_{\epsilon}(\mathcal{X}^2) \; \mbox{ for some } \epsilon \geq 0.
\eq
If $H_0$ is true, $H_0'$ is also true but the converse may not be true even though the probability of the converse not true would be relatively small. The statistics for testing $H_0'$ need to account for multiple comparisons over filtration values $\epsilon \geq 0$. As a statistic for testing $H_0'$, $D_q$ was used in discriminating brain networks \cite{chung.2013.MICCAI,lee.2017.HBM,chung.2016.twin}. The test statistic takes care of the multiple comparisons by the use of supremum.  

One can also test if the change of $\beta_0$ over the filtration are different between two networks using the integral statistic approach \cite{chung.2013.CNA}. We test 
\bq H_0'': \int_{0}^{\infty} f \circ \mathcal{B}_{\epsilon}(\mathcal{X}^1)  \;d\epsilon &=& \int_{0}^{\infty} f \circ \mathcal{B}_{\epsilon}(\mathcal{X}^2) \; d\epsilon\\
& vs. &\\
H_1'': \int_{0}^{\infty} f \circ \mathcal{B}_{\epsilon}(\mathcal{X}^1) \;d\epsilon &=& \int_{0}^{\infty} f \circ \mathcal{B}_{\epsilon}(\mathcal{X}^2) \; d\epsilon.\eq
Obviously, the integral statistic of the from 
$$ \int_{0}^{\infty} \left| f \circ \mathcal{B}_{\epsilon}(\mathcal{X}^1) -  f \circ \mathcal{B}_{\epsilon}(\mathcal{X}^1) \right|  \;d\epsilon$$
can be used to test $H_0''$. By taking the integral over all filtration values $\epsilon$, we are also accounting for the  multiple comparisons. 
The probability distribution of the integral statistic under $H_0''$ can be obtained by resampling techniques such as Jackknife \cite{chung.2015.TMI} or permutation test \cite{lee.2017.HBM}. On the other hand, due to the simplicity of $D_q$, we can drive the probability distribution of $D_q$ exactly under $H_0'$.

\begin{figure}[t]
\centering
\includegraphics[width=1\linewidth]{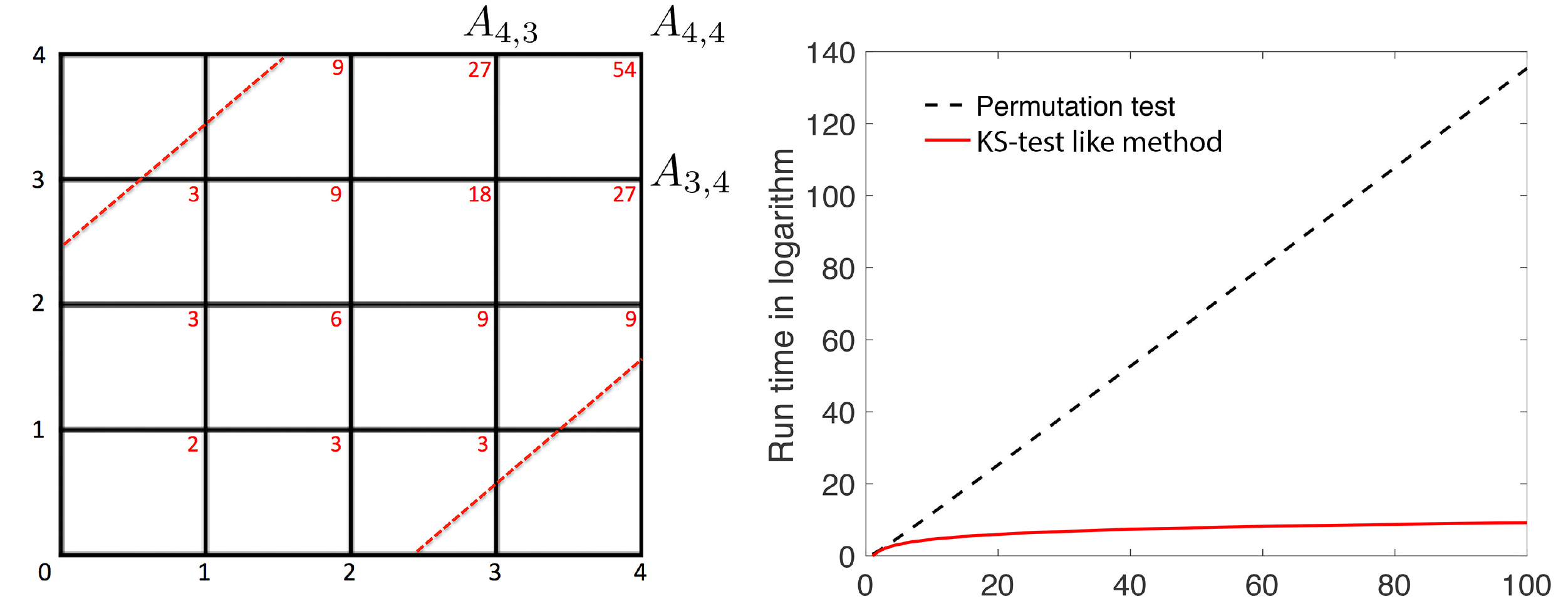} 
\caption{Left: $A_{u,v}$ are computed within the boundary (dotted red line). The red numbers computed $A_{u,v}$. The computation is done from $(0,0)$ to $(u,v)$ in an iterative fashion. Right:}
\label{fig:App}
\end{figure}

\begin{theorem}
\label{theorem:lim1}
$$P (D_q \geq d )   = 1 - \frac{A_{q,q}}{{2q \choose q}},$$
where $A_{u,v}$ satisfies $A_{u,v} = A_{u-1,v} + A_{u, v-1}$
with the boundary condition $A_{0,q}=A_{q,0}=1$ within band $|u - v| < d$. 
\end{theorem}

The proof is given in  \cite{chung.2016.twin} and it is based on the combinatorial emulation of all possible permutations satisfying condition $D_q \geq d$ in the sample space when we permute the two monotonic feature vectors. Theorem \ref{theorem:lim1} provides the exact probability computation.

\begin{example} Probability $P(D_4 \geq 2.5)$ is computed iteratively as follows. We start with computing 
\bq A_{1,1} &=& A_{0,1} + A_{1,0} =2,\\
 A_{2,1} &=& A_{1,1} + A_{1,0} = 3, \\
& \vdots & \\
A_{4,4} &=& A_{4,3}+ A_{3,4} = 27+ 27 = 54.\eq
The red numbers in Fig. \ref{fig:App}-left are iteratively enumerated $A_{u,v}$. The probability is computed as $$P(D_4 \geq 2.5) = 1- 54/ {8 \choose 4}=0.23.$$
\end{example}

Computing $A_{q,q}$ iteratively requires at most $q^2$ operations while permuting two samples consisting of q elements each requires ${2q \choose q}$ operations. Thus, our method can compute the $p$-value substantially faster than the permutation test that is exponentially slow. The asymptotic probability distribution of $D_q$ can be also determined for sufficiently large $q$ without computing iteratively as in Theorem \ref{theorem:lim1}.

\begin{theorem} 
\label{theorem:lim2}
$\lim_{q \to \infty}  \Big( D_q /\sqrt{2q} \geq  d  \Big)  = 2 \sum_{i=1}^{\infty} (-1)^{i-1}e^{-2i^2d^2}.$
\end{theorem}
The result follows from \cite{gibbons.1992,smirnov.1939}. From Theorem \ref{theorem:lim2}, $p$-value under $H_0'$ is computed as
$$\mbox{$p$-value} = 2 e^{-d_{o}^2} - 2e^{-8d_{o}^2} + 2 e^{-18d_{o}^2} \cdots,$$ where $d_{o}$ is the observed value of $D_{q}/\sqrt{2q}$ in the data.  For any large observed value $d_0 \geq 2$, the second term is in the order of $10^{-14}$ and insignificant. Even for small observed $d_0$, the expansion converges quickly and 5 terms are sufficient.\\

\begin{figure}[t]
\centering
\includegraphics[width=1\linewidth]{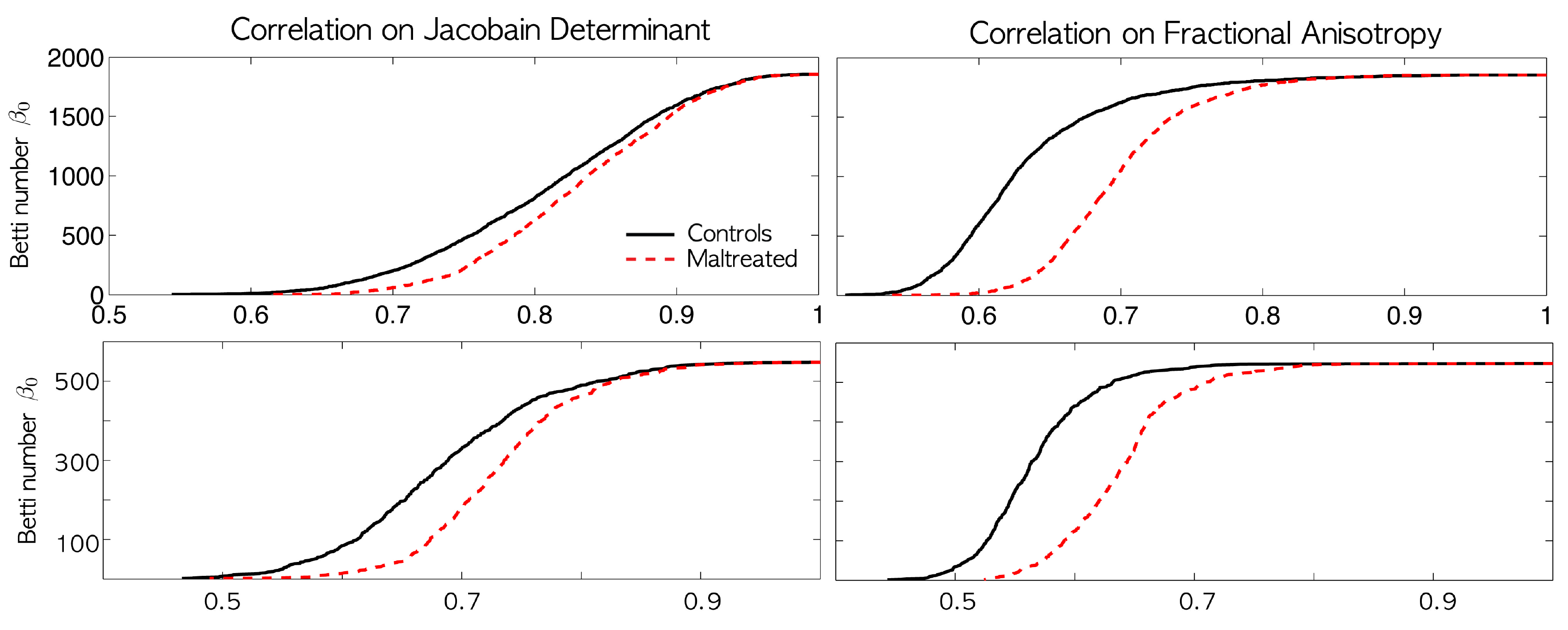}
\caption{$\beta_0$-plots over 1 - correlation showing structural network differences between maltreated children (dotted red) and normal controls (solid black) \cite{chung.2015.TMI}.  1856 (top) and 548 (bottom) node networks show the similar patterns of group differences.}
\label{fig:KS-example}
\end{figure}

\begin{example} The KS-test like distance is applied to multimodal magnetic resonance imaging (MRI) and diffusion tensor imaging (DTI) study of 31 normal controls and 23 age-matched children who experienced maltreatment while living in post-institutional settings. The study detail is given in \cite{chung.2013.MICCAI,chung.2015.TMI}. 548 and 1856 nodes are uniformly sampled along the white matter boundary of the brain. The correlations of the Jacobian determinant for MRI and fractional anisotropy (FA) values for DTI were computed between the nodes. 1- correlations were used as edge weights. $\beta_0$-plots visually show the group separation (Fig. \ref{fig:KS-example}). For Jacobian determinant, the observed $d$ is 252. For FA-value, the observed $d$ is 766. In both cases, $p$-values are less than 0.001 showing strong group differences. 
 \end{example}

There is no known method for computing the exact probability distributions for distance measures other than the KS-test like distance. Unless we are dealing with significantly large number of observed distances and willing to resort to the asymptotic normality based on the central limit theorem \cite{bubenik.2015},  it is necessary to use  resampling techniques in estimating the null distributions. 

\section{Network resampling}

Statistical inference on network distances can be done using the permutation test or bootstrap \cite{chung.2013.MICCAI,efron.1982,lee.2012.TMI}. Here we explain the permutation test procedure for network distances. The usual setting in brain imaging applications is a two-sample comparison. 
Suppose there are $m$ measurement in Group 1 on node set $V$ of size $p$. Denote the data matrix as ${\bf X}^1_{m \times p}$. The edge weights of Group 1 is given by  $f({\bf X}^1)$  for some function $f$ and the metric space is given by $\mathcal{X}^1 = (V, f({\bf X}^1))$. Suppose there are $n$ measurement in Group 2 on the identical node set $V$. Denote data matrix as ${\bf X}^2_{n \times p}$ and the corresponding metric space as $\mathcal{X}^1 = (V, f({\bf X}^1))$.  We test the statistical significance of  network distance $D(\mathcal{X}^1, \mathcal{X}^2)$ under the null hypothesis $H_0$. 

The permutations are done as follows. Concatenate the data matrices
$${\bf X} = (x_{ij}) =
\left(
\begin{array}{c}
  {\bf X}^1     \\
  {\bf X}^2   \\
\end{array}
\right)_{(m+n) \times p}.$$
Now permute the indices of row vectors of ${\bf X}$ in the symmetric group of degree $m+n$, i.e., $S_{m+n}$ \cite{kondor.2007}. The permuted data matrix is denoted as ${\bf X}_{\sigma(i)} = (x_{\sigma(i), j})$, where $\sigma \in S_{m+n}$. Then we split ${\bf X}_{\sigma(i)}$ into submatrices such that
$${\bf X}_{\sigma(i)} =
\left(
\begin{array}{c}
  {\bf X}_{\sigma(i)} ^1     \\
  {\bf X}_{\sigma(i)} ^2   \\
\end{array}
\right),$$
where ${\bf X}_{\sigma(i)} ^1$ and  ${\bf X}_{\sigma(i)} ^2$ are of sizes $m \times p$ and $n \times p$ respectively.  Let $\mathcal{X}^1_{{\sigma}(i)} = (V, f({\bf X}^1_{{\sigma}(i)} ))$ and $\mathcal{X}^2_{{\sigma}(i)} = (V, f({\bf X}^2_{{\sigma}(i)} ))$ be weighted networks where the rows of the data matrices are permuted across the groups. Then we have distance $D(\mathcal{X}^1_{{\sigma}(i)} , \mathcal{X}^2_{{\sigma}(i)} )$ for each permutation. The number of permutations exponentially increases and it is impractical to generate every possible permutation. So up to tens of thousands permutations are generated in practice. This is an approximate method and a care should be taken to guarantee the convergence.

\section{Discussion}

{\em Different node sizes.}  Even if the size of node sets differ in two networks, network distances  can be still computed. Consider two networks $(V_1, \rho_{1})$ and $(V_2, \rho_{2})$. If $V_1 \neq V_2$, we will increase the size of the node sets in the two networks to the larger node set $V_1 \cup V_2$. The metric $\rho_1$ will be changed to the new metric $\tilde \rho_1$, which is defined as follows. $\tilde \rho_1$ restricted to the original smaller node set $V_1$ will be $\rho_1$, i.e., $\tilde \rho_1 | V_1 = \rho_1$. All other entries of $\tilde \rho_1$ will be zero. We extend the metric $\rho_2$ to $\tilde \rho_2$ similarly. Then new networks $(V_1 \cup V_2, \tilde \rho_1)$ and $(V_1 \cup V_2, \tilde \rho_2)$ will have the identical node set. \\

\begin{wrapfigure}{r}{0.5\textwidth}
\vspace{-0.5cm}
\centering
\includegraphics[width=1\linewidth]{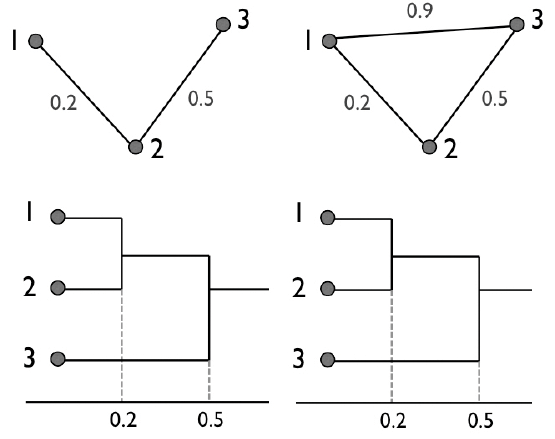}
\caption{Two topologically distinct graphs may have identical dendrograms. This shows the dendrogram and  GH distance. 
The dendrogram  is equivalent to the construction of the minimum spanning tree (MST) of a network, which is basically a tree. Dendrograms cannot properly represent cycles and loops in networks.}
\label{fig:topology-dendro2}
\vspace{-0.5cm}
\end{wrapfigure}

\noindent{\em The limitation of GH distance.}  The limitation of the single linkage matrix is the inability to discriminate a cycle in a graph. Consider two topologically different graphs with three nodes (Fig. \ref{fig:topology-dendro2}). 
However, the corresponding single linkage matrices are identically given by 
$$
 \left( \begin{array}{ccc}
0 & 0.2 & 0.5\\ 
0.2 & 0 & 0.5\\
 0.5    &0.5  & 0
\end{array} \right) \mbox{ and }  \left( \begin{array}{ccc}
0 & 0.2 & 0.5\\ 
0.2 & 0 & 0.5\\
 0.5    &0.5  & 0
\end{array} \right).$$
The lack of uniqueness of SLMs makes GH distance incapable of discriminating networks with cycles. 
The single linkage matrices can be further used in obtaining the dendrogram representation of the brain networks. Thus, dendrograms also have difficulty representing and discriminating the cycles in the network. The dendrogram representation is based on the minimum spanning tree (MST) of a graph, which is basically a tree presentation \cite{lee.2011.MICCAI,lee.2012.TMI}. Hence dendrogram cannot properly represent cycles  in the network.\\

\noindent{\em Exact topological inference.} 
Theorem \ref{theorem:lim1} is the exact nonparametric procedure and does not assume any statistical distribution on graph features other than that they has to be monotonic. The technique is very general and applicable to other monotonic graph features such as node degree. 

Based on Stirling's formula, $$q! \sim \sqrt{2\pi q} \Big(\frac{q}{e}\Big)^q,$$ the total number of permutations in permuting two vectors of size $q$ each is $${2q \choose q} \sim \frac{4^q}{\sqrt{2\pi q}}.$$ 
This is substantially larger than the quadratic run time $q^2$ needed in our method (Fig. \ref{fig:App}-right). Even for small $q=10$, more than tens of thousands permutations are needed for the accurate estimation the $p$-value. On the other hand, only up to $100$ iterations are needed in the KS-test like distance. It is likely that the method can be extended to much wider class of graph and persistent homology features. This is a left as a future study.

\section*{Acknowledgements}
We would like to thank Drs.  Richard Davidson, Seth Pollack of University of Wisconsin-Madison and Dong Soo Lee of Seoul National University Hospital for brain imaging data used in the examples. We would also like to thank Victor Solo for pointing error related to Example 1. 

This work is supported by NIH Brain Initiative Award R01 EB022856 and Basic Science Research Program through the National Research Foundation (NRF) of Korea (NRF-2016R1D1A1B03935463).

\bibliographystyle{plain}
\bibliography{reference.2016.12.31,leehk,bib_yuan}

\end{document}